\documentclass[aps,prl,twocolumn,superscriptaddress,prbib]{revtex4}

\usepackage{graphicx}%
\usepackage{dcolumn}
\usepackage{amsmath}
\usepackage{color}
\usepackage{multirow}
\usepackage{amssymb}

\makeatletter
\def\btt#1{\texttt{\@backslashchar#1}}%
\DeclareRobustCommand\bblash{\btt{\@backslashchar}}%
\makeatother

\begin{document}

\preprint{PREPRINT (\today)}

\title{Isotropic single gap  superconductivity of elemental Pb: the `smiling' approach}

\author{Rustem Khasanov}
\email{rustem.khasanov@psi.ch}
  \affiliation{Laboratory for Muon Spin Spectroscopy, Paul Scherrer Institut, CH-5232 Villigen PSI, Switzerland}

\author{Debarchan Das}
  \affiliation{Laboratory for Muon Spin Spectroscopy, Paul Scherrer Institut, CH-5232 Villigen PSI, Switzerland}

\author{Dariusz Jakub Gawryluk}
  \affiliation{Laboratory for Multiscale Materials Experiments, Paul Scherrer Institut, Villigen CH-5232, Switzerland}

\author{Ritu Gupta}
  \affiliation{Laboratory for Muon Spin Spectroscopy, Paul Scherrer Institut, CH-5232 Villigen PSI, Switzerland}

\author{Charles Mielke III}
  \affiliation{Laboratory for Muon Spin Spectroscopy, Paul Scherrer Institut, CH-5232 Villigen PSI, Switzerland}


\begin{abstract}

The unconventional multi-gap superconductivity in elemental Pb were reported previously by surface sensitive tunneling experiments, as well as predicted by several theory works.
To obtain bulk  evidence for such multiple gap behavior, the thermodynamic critical field $B_{\rm c}$ was measured along three different crystallographic directions ([100], [110], and [111]) in a high-quality Pb single crystal  by means of muon spin rotation/relaxation. No difference in temperature evolution of $B_{\rm c}$ for all three directions was detected. The average reduced gap $\alpha=\Delta/k_{\rm B}T_{\rm c}=2.312(3)$ ($\Delta$ is the zero-temperature gap value and $T_{\rm c}$ is the transition temperature) was further obtained by employing the phenomenological $\alpha-$model.
Our results imply that the elemental Pb is an {\it isotropic} superconductor with a {\it single} energy gap.
\end{abstract}

\maketitle



Superconductivity was first discovered by Kamerlingh Onnes in elemental Hg in 1911  \cite{Kammerlingh_Hg_1911}, and then in elemental Sn and Pb within the next two years  \cite{Kammerlingh_Hg_1913}. To date, 31 elements are known to be superconducting at ambient pressure \cite{Superconductors_wikipedia} and, among them, Pb has the second highest superconducting transition temperature $T_{\rm c}\simeq 7.20$~K, just after elemental Nb with $T_{\rm c}\simeq 9.2$~K.

The description of lead within the weak-coupled BCS formalism was unsuccessful, thus leading to the corresponding development of the strong-coupled extension within the framework of the Eliashberg theory (see Ref.~\onlinecite{ Carbotte_RMP_1990} and references therein). The `unconventional' aspects of superconductivity in Pb, in relation to the `conventional' BCS ones, were discussed in a series of papers \cite{Bennett_PR_1965, Tomlinson_PRB_1976, VanDerHoeven_PR_1965, Overhauser_PRL_1988, Carbotte_RMP_1990, Short_PRL_2000, Blackford_PR_1969,Lykken_PRB_1971,Ruby_PRL_2015, Floris_PRB_2007,Saunderson_PRB_2020}. In particular, the presence of at least two separate superconducting energy gaps with distinct values were reported experimentally \cite{Blackford_PR_1969,Lykken_PRB_1971,Ruby_PRL_2015}, as well as proposed theoretically \cite{Floris_PRB_2007,Saunderson_PRB_2020}.

Evidence for multi-gap superconductivity should be present in the temperature evolution of various thermodynamic quantities, {\it e.g.}, electronic specific heat, entropy, critical fields, {\it etc.}. Among them, the thermodynamic critical field $B_{\rm c}$, which is related to the condensation energy of superconducting carriers via \cite{Tinkham_book_1975}:
\begin{equation}
\frac{B_{\rm c}^2(0)}{4\pi} = N_{\rm F}(0)\Delta^2
 \label{eq:condensation-energy}
\end{equation}
[$B_{\rm c}(0)=B_{\rm c}(T=0)$, $N_{\rm F}(0)$ is the density of states at the Fermi level, and $\Delta$ is the zero-temperature value of the superconducting energy gap], becomes a direct probe  of multiple superconducting energy gaps \cite{Zehetmayer_JLTP_2003, Nicol_PRB_2005, Khasanov_AuBe_PRR_2020, Khasanov_AuBe_PRB_2020}, as well as allows one to probe the superconducting gap anisotropy \cite{Clem_AnnPhys_1966, Clem_PR_1967, Khasanov_Al_PRB_2021}. As the temperature dependence of $B_{\rm c}$ normalized
to its zero-temperature value follows very closely a nearly
quadratic behavior, the deviation function $D(T)$ is normally considered:
\begin{equation}
D(T)=\frac{B_{\rm c}(T)}{B_{\rm c}(0)} - \left[1-\left(\frac{T}{T_{\rm c}}\right)^2\right].
 \label{eq:Deviation-function}
\end{equation}

\begin{figure}[tbh]
\includegraphics[width=1.0\linewidth]{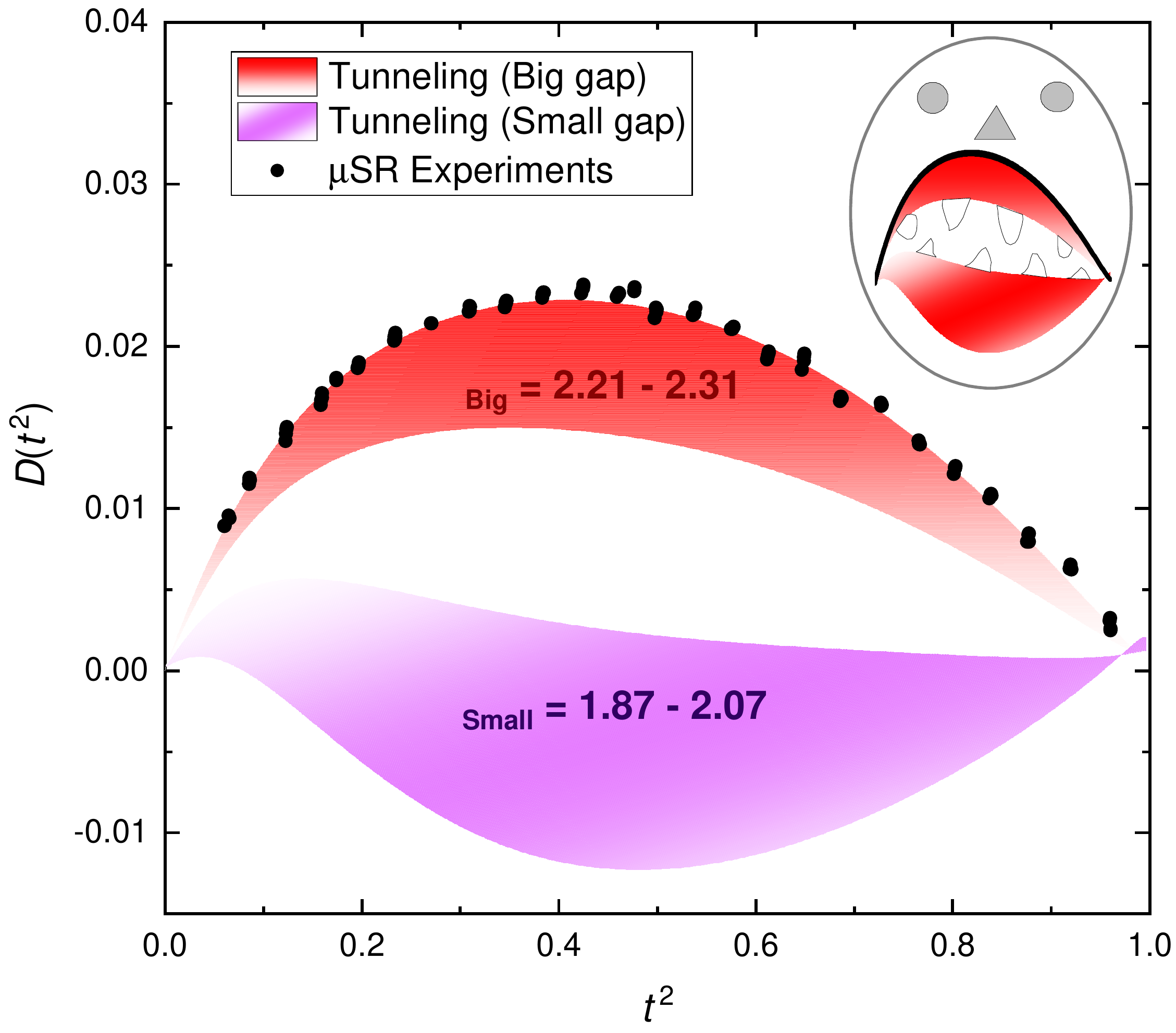}
\caption{Deviation of the thermodynamic critical field $B_{\rm c}(T)$ from the parabolic behavior [$D(t^2)=B_{\rm c}(t^2)/B_{\rm c}(0)-(1-t^2)$] as expected for the `big': $\alpha_{\rm big}=\Delta_{\rm big}/k_{\rm B}T_{\rm c}=2.21-2.31$ (red segment) and  the `small': $\alpha_{\rm small}=\Delta_{\rm small}/k_{\rm B}T_{\rm c}=1.87-2.07$ (violet segment) reduced gaps for elemental Pb reported in Refs.~\onlinecite{Blackford_PR_1969,Lykken_PRB_1971,Ruby_PRL_2015} ($t=T/T_{\rm c}$ is the reduced temperature).
Calculations of $D(t)$ functions are performed by using the $\alpha-$model \cite{Padamsee_JLTP_1973, Johnston_SST_2013}. Black points are the experimental $D(t)$ data measured with the external field $B_{\rm ex}$ applied along three different crystallographic directions: [100], [110], and [111] (see text for details). The picture at the top right corner is the `smiling lead' with the `lips' formed by simulated $D(t)$ areas and `mustache' as obtained from the present  experiment.  }
\label{fig:Deviation-Tunneling}
\end{figure}

By combining experimental data of Refs.~\onlinecite{Blackford_PR_1969,Lykken_PRB_1971,Ruby_PRL_2015}, the `big' and the `small' reduced gaps  in Pb were found to stay in the range of $\alpha_{\rm big}=\Delta_{\rm big}/k_{\rm B}T_{\rm c}=2.21-2.31$ and $\alpha_{\rm small}=\Delta_{\rm small}/k_{\rm B}T_{\rm c}=1.87-2.07$. The corresponding  deviation functions, calculated by employing the $\alpha-$model of Padamsee {\it et al.} \cite{Padamsee_JLTP_1973, Johnston_SST_2013}), are shown in Fig.~\ref{fig:Deviation-Tunneling}. The top and bottom branches of the $D(t)$ distributions ($t=T/T_{\rm c}$ is the reduced temperature) represent kind of  `upper' and `lower' lips which, when embedded into an oval, form the image of a `smiling lead' (right top corner of Fig.~\ref{fig:Deviation-Tunneling}).
It should be stressed, however, that experiments pointing to the presence of two distinct superconducting energy gaps in elemental Pb have been performed so far by means of tunneling only, {\it i.e.} by a surface sensitive technique. It is not obvious if the band structure and thus the gap functions would remain the same in the bulk as at the surface. Note that disagreement between theory and tunneling data have already been mentioned by Saunderson {\it et al.} in Ref.~\onlinecite{Saunderson_PRB_2020}.

In this work, in order to test whether the superconducting gap structure of elemental lead is of a single- or a two-gap type, measurements of the thermodynamic critical field $B_{\rm c}(T)$ along three crystallographic directions were performed by means of the muon-spin rotation/relaxation ($\mu$SR) technique.
Experiments were carried out  by applying external magnetic field $B_{\rm ex}$ along [100], [110], and [111] axes.  The selection of these orientations comes from the face-centered cubic (fcc) crystal structure of Pb, where the strongest differences in physical quantities are expected while measuring along the main
axis and two diagonals. The orientational dependences of the electron-phonon coupling constant, the superconducting energy gaps and densities of states at the Fermi level \cite{Blackford_PR_1969,Lykken_PRB_1971,Ruby_PRL_2015, Floris_PRB_2007,Saunderson_PRB_2020} are expected to affect the shape of $B_{\rm c}(T)$ curves, as well as to change the absolute value of $B_{\rm c}(0)$.  In our experiments, however, all measured $B_{\rm c}(T)$ dependences were found to coincide within the experimental accuracy. The corresponding deviation functions $D(t)$'s lie just at the top border of the `big' superconducting energy gap branch and forms a `mustache' on the `smiling lead' face (see Fig.~\ref{fig:Deviation-Tunneling}). Our results suggest that elemental Pb is an {\it isotropic} and a {\it single-gap} superconductor.

\begin{figure}[tbh]
\includegraphics[width=0.95\linewidth]{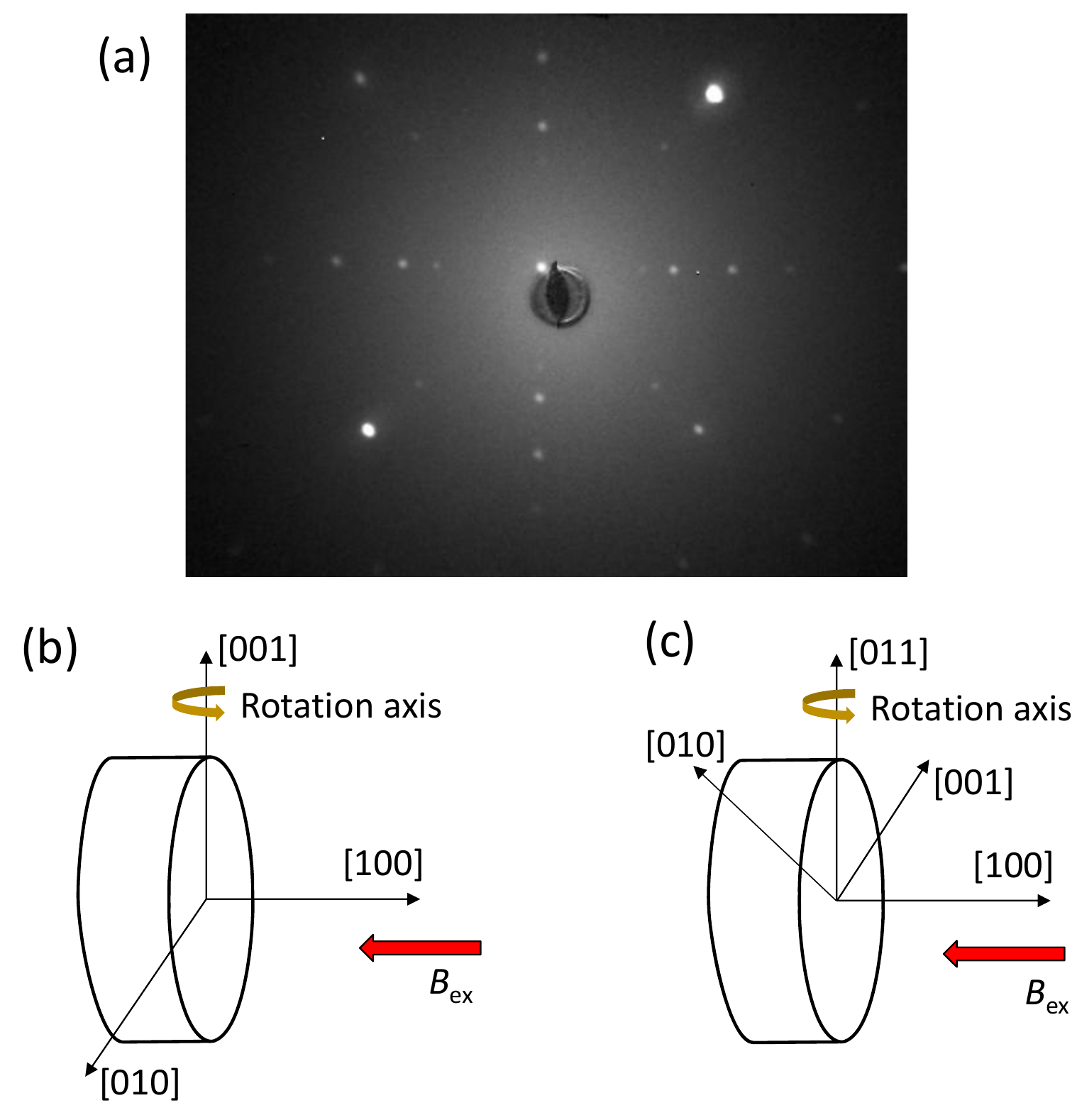}
\caption{(a) Laue image of Pb single crystal. The cubic crystal structure of Pb is clearly visible. (b) The schematic representation of the first set of experiments (exp.\#1) allowing sample rotation along 001 crystal axis. In this configuration $B_{\rm ex} \parallel [100]$ and $B_{\rm ex} \parallel [110]$ sets of measurements were performed. (c) The second set of experiments (exp.\#2) with the sample rotation along 011 axis. In these experiments, $B_{\rm ex} \parallel [100]$ and $B_{\rm ex} \parallel [111]$ configurations were probed. }
\label{fig:Sample-Laue_Rotation}
\end{figure}

The Pb single crystal sample was supplied by Goodfellow \cite{Goodfellow_Pb}. The disk shaped single crystal has the following parameters: orientation -- [100], thickness -- 1.6~mm, diameter -- 10~mm, and purity -- 99.999\%. Prior to $\mu$SR studies, in order to minimize pinning effects, the sample was annealed for one week at a temperature of $300^{\rm o}$C, {\it i.e.}, $\simeq 27^{\rm o}$C below the melting point. The Laue image of the Pb single crystal is presented in Fig.~\ref{fig:Sample-Laue_Rotation}~(a), which clearly evinces the cubic crystal structure of Pb.

The transverse-field  muon-spin rotation/relaxation (TF-$\mu$SR) experiments were conducted using the  Dolly spectrometer ($\pi$E1 beam line) at the Paul Scherrer Institute, Switzerland. Measurements were performed in the intermediate state of superconducting Pb, {\it i.e.}, when the
sample volume is separated into the normal state and the superconducting state
(Meissner) domains \cite{Tinkham_book_1975, Khasanov_Al_PRB_2021, Khasanov_AuBe_PRR_2020, Kittel_book_2007, deGennes_book_1966, Prozorov_PRL_2007, Prozorov_NatPhys_2008, Khasanov_Bi_PRB_2019, Karl_2019_PRB, Khasanov_Ga_PRB_2020}. Two sets of experiments were performed. In the first experiment [exp.\#1, Fig.~\ref{fig:Sample-Laue_Rotation}(b)], the sample was rotated along the 001 axis, which allowed for measurements with the external magnetic field $B_{\rm ex}$ applied parallel to [100] ($B_{\rm ex} \parallel [100]$) and [110] ($B_{\rm ex} \parallel [110]$) axes.  In the second experiment [exp.\#2, Fig.~\ref{fig:Sample-Laue_Rotation}(c)], the sample was rotated along [011] axis, so the corresponding $B_{\rm ex} \parallel [100]$ and $B_{\rm ex} \parallel [111]$  measurements were made.

The measurement procedure was as follows. First, a predetermined temperature below the superconducting transition temperature [$T_{\rm c}(B_{\rm ex}=0)\simeq 7.2$~K] was stabilized.  Second, the sample was turned into the $B_{\rm ex} \parallel [100]$ configuration. Then, $B_{\rm ex}$ was increased up to 85~mT, {\it i.e.} above $B_{\rm c}(0)\simeq 80$~mT \cite{Kittel_book_2007, Chanin_PRB_1972, Brandt_ZhETF_1975}. Finally, the TF-$\mu$SR measurements were performed at fields corresponding to $\simeq 90$, 85, 80, and 75\% of $B_{\rm c} (T)$, by considering the $B_{\rm c}(T)$ curve determined in Refs.~\onlinecite{Chanin_PRB_1972, Brandt_ZhETF_1975}. By finishing experiments in the $B_{\rm ex} \parallel [100]$ configuration and by keeping the  temperature unchanged, the sample was rotated by $45^{\rm o}$. After such rotation, measurements at similar fields were repeated for $B_{\rm ex} \parallel [110]$ [exp.\#1, Fig.~\ref{fig:Sample-Laue_Rotation}(b)] and for $B_{\rm ex} \parallel [111]$ [exp.\#2, Fig.~\ref{fig:Sample-Laue_Rotation}(c)].
Note that the experiments with the sample rotation at constant $T$, compared to $T-$scan at constant angle, ensures that the sample temperature remains the same for two different field orientations. In this case, possible differences in the measured values of the thermodynamic critical fields are caused by intrinsic orientational effects and not by the possible temperature differences/instabilities. This also implies that measurements in the $B_{\rm ex} \parallel [100]$ configuration were performed twice: once in exp.\#1 and a second time in exp.\#2.
%


The magnetic field distribution in a type-I superconductor in the intermediate state, which is probed directly by means of TF-$\mu$SR, consists of two sharp
peaks corresponding to the response of the domains remaining in the Meissner state ($B = 0$) and in the normal state ($B\equiv B_{\rm c} > B_{\rm ex}$). Consequently, the value of $B_{\rm c}$ could be directly and very precisely determined by measuring the position of the $B > B_{\rm ex}$ peak \cite{Khasanov_Al_PRB_2021, Khasanov_AuBe_PRR_2020, Kittel_book_2007, deGennes_book_1966, Prozorov_PRL_2007, Prozorov_NatPhys_2008, Khasanov_Bi_PRB_2019, Karl_2019_PRB, Khasanov_Ga_PRB_2020, Egorov_PRB_64, Gladisch_HypInt_1979, Grebinnik_ZhETF, Leng_PRB_2019, Beare_PRB_2019, Kozhevnikov_JSNM_2020}. The details of the TF-$\mu$SR experiments on elemental Pb and the data analysis procedure are discussed in the Supplemental Material \cite{SupplementalMaterial}.

\begin{figure}[tbh]
\includegraphics[width=0.95\linewidth]{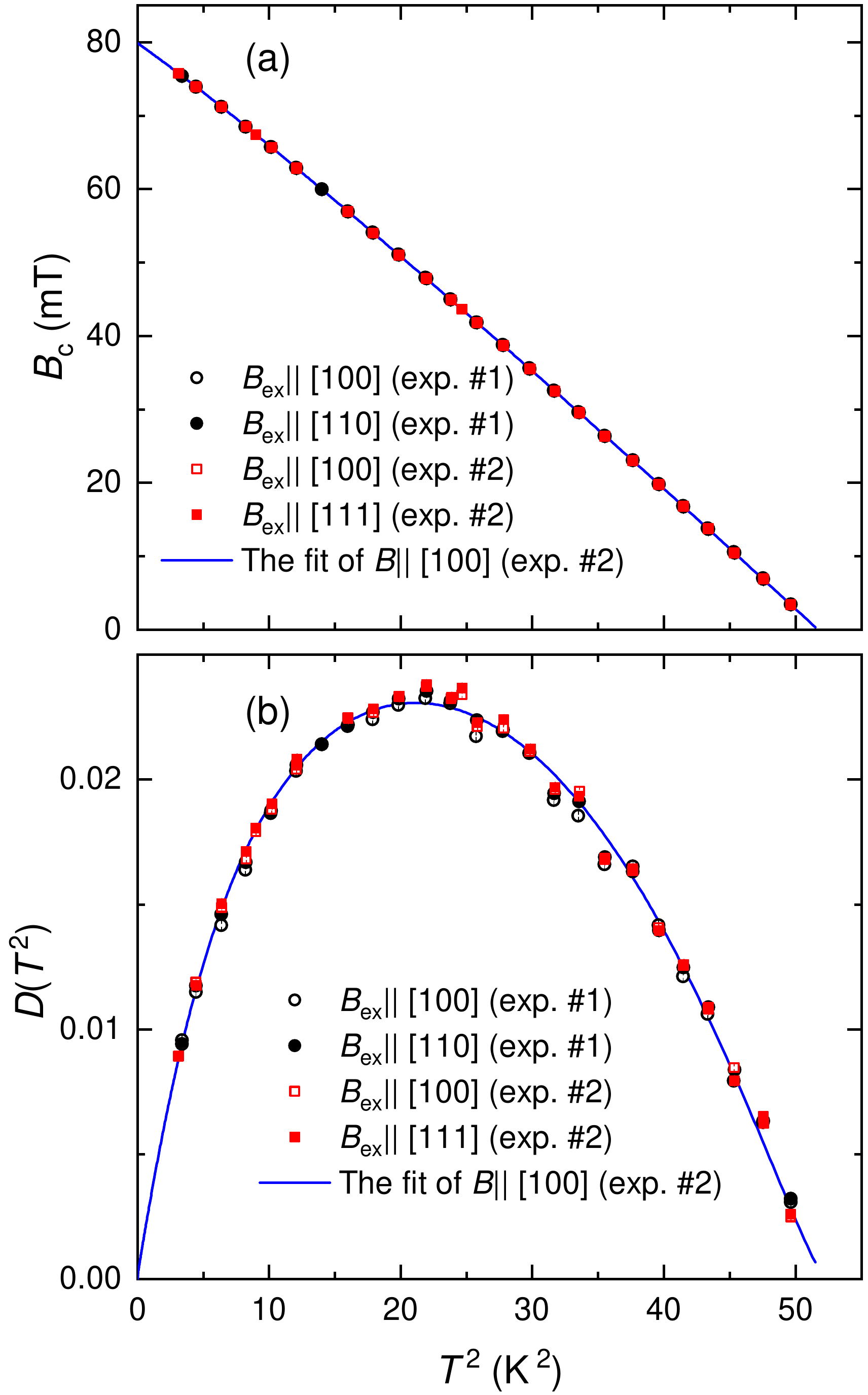}
\caption{(a) Temperature dependences of the thermodynamic
critical field $B_{\rm c}$ of elemental Pb single crystal measured along different crystallographic directions. The solid line is the fit of $B_{\rm c}(T^2)$ measured in exp.\#2 in $B_{\rm ex}\parallel [100]$ configuration by means of Eq.~\ref{eq:alpha-model_Padamsee}. (b) Deviation functions $D(T^2)$ obtained by subtracting parabolic functions from measured $B_{\rm c}(T^2)$ curves, see Eq~\ref{eq:Deviation-function}. The solid line is the same as in panel (a), but after subtracting the parabolic function.}
\label{fig:Bc}
\end{figure}

The results of TF-$\mu$SR experiments are summarized in Figure~\ref{fig:Bc}. Panels (a) and (b) represent the temperature evolution of the thermodynamic critical field $B_{\rm c}$ and the deviation function $D$, respectively. Note that within each set of data [$B_{\rm c}(T^2)$ or $D(T^2)$], the experimental dependences measured along three different crystallographic directions ([100], [110], and [111]) cannot be distinguished from each other. The individual $B_{\rm c}(T^2)$ and $D(T^2)$ curves are presented in the Supplemental Material \cite{SupplementalMaterial}.

Bearing in mind that in elemental Pb two superconducting energy gaps might be present (see Refs.~\onlinecite{Blackford_PR_1969,Lykken_PRB_1971,Ruby_PRL_2015, Floris_PRB_2007,Saunderson_PRB_2020} and Fig.~\ref{fig:Deviation-Tunneling}), the measured $B_{\rm c}(T)$ dependences should be analyzed within the multi-gap scenario. Such approach was recently used by us to describe the two-gap superconductivity in noncentrosymmetric binary alloy AuBe \cite{Khasanov_AuBe_PRR_2020, Khasanov_AuBe_PRB_2020}. However, as will be shown later, in elemental Pb, the single-gap approach was found to describe the experimental data reasonably precisely, so any type of admixture of the second superconducting energy gap was not expected [see also the single-gap theory curves in panels (a) and (b) of Fig.~\ref{fig:Bc}].

The $B_{\rm c}(T^2)$ curves were analyzed within the framework of the empirical $\alpha-$model of Padamsee {\it et al.} \cite{Padamsee_JLTP_1973, Johnston_SST_2013}. Here, the version for strong-coupled superconductors, such as elemental Pb, was employed. It accounts for the fact that in metals with strong electron-phonon coupling, the electronic specific heat coefficient $\gamma_{\rm e}$  is significantly temperature dependent. Consequently, the basic assumptions of the weak-coupled $\alpha-$model:
\begin{equation}
C_{\rm en}(T)= S_{\rm en}(T)=\gamma_{\rm e}T
 \label{eq:weak-coupled_specific-heat}
\end{equation}
must be substituted with \cite{Grimvall_PhConMat_1969, Giustino_RMP_2017}:
\begin{equation}
C_{\rm en}(T)=T\frac{\partial S_{\rm en}(T)}{\partial T}=\gamma_0 \left[ 1+\lambda_{\rm el-ph}\frac{\gamma_1(T)}{\gamma_1(0)} \right] T.
 \label{eq:strong-coupled_specific-heat}
\end{equation}
Here, $C_{\rm en}$ and  $S_{\rm en}$ are the normal state electronic specific heat and the electronic entropy, respectively; $\lambda_{\rm el-ph}$ is the electron-phonon coupling constant; $\gamma_1(T)$ is the function accounting for nonlinear temperature dependence of the electronic specific heat; and $\gamma_0$ is the electronic specific heat coefficient in the absence of electron-phonon coupling. Note that by setting $\lambda_{\rm el-ph}=0$, {\it i.e.} by ignoring the electron-phonon interaction, Eq.~\ref{eq:strong-coupled_specific-heat} reduces to Eq.\ref{eq:weak-coupled_specific-heat}. For elemental Pb, the quantity $\gamma_1(T)/\gamma_1(0)$ was calculated by Grimvall \cite{Grimvall_PhConMat_1969}, and it is presented in Fig.~S6 in the Supplemental Material \cite{SupplementalMaterial}.

Within the strong-coupling approach, the thermodynamic critical field is further obtained as \cite{Padamsee_JLTP_1973, Johnston_SST_2013}:
\begin{equation}
\frac{B_{\rm c}^2(t)}{8\pi S_{\rm en}(1)} = \int_t^{1}\frac{S_{\rm en}(t')}{S_{\rm en}(1)}dt'-\int_t^{1}\frac{S_{\rm es}(t')}{S_{\rm es}(1)}dt',
 \label{eq:alpha-model_Padamsee}
\end{equation}
with the temperature dependences of the  normal state $S_{\rm en}(t)$ and the superconducting state $S_{\rm es}(t)$ entropies described as \cite{Padamsee_JLTP_1973, Johnston_SST_2013}:
\begin{equation}
S_{\rm en}(t)= \gamma_0 \int_0^t \left[ 1+\lambda_{\rm el-ph}\frac{\gamma_1(t')}{\gamma_1(0)} \right] dt' \nonumber
\end{equation}
and
\begin{equation}
S_{\rm es}(t) = \frac{6\alpha^2 S_{\rm en}(1)}{\pi^2 t}\int_0^\infty f(\alpha, E, t)
\left( E+\frac{\varepsilon^2}{E} \right) d\varepsilon. \nonumber
\end{equation}
Here, $t=T/T_{\rm c}$ is the reduced temperature, $\alpha=\Delta/k_{\rm B} T_{\rm c}$, $f(\alpha,E,t) = [\exp(\alpha E/t)+1]^{-1}$ is the Fermi function, and $E=E[\varepsilon,\Delta(t)/\Delta]=\sqrt{\varepsilon^2+[\Delta(t)/\Delta]^2}$ is the quasiparticle energy. The
temperature dependence of the normalized gap, tabulated by
M\"{u}hlschlegel \cite{Muehlschlegel_ZPhys_1959} was parametrized as $\Delta(t)/\Delta = \tanh\{1.82[1.018(1/t-1)]^{0.51}\}$ \cite{Khasanov_AuBe_PRB_2020}.

\begin{table}[htb]
\caption{\label{Table1} The parameters obtained from the analysis of measured $B_{\rm c}(T)$ dependences in Pb single crystal within the framework of a strong-coupling $\alpha-$model \cite{Padamsee_JLTP_1973, Johnston_SST_2013}. The meaning of the parameters is the following: $T_{\rm c}$ is the superconducting transition temperature, $B_{\rm c}(0)$ is the zero-temperature value of the upper critical field, $\alpha=\Delta/k_{\rm B}T_{\rm c}$ is the reduced gap, and $\Delta$ is the zero-temperature value of the superconducting energy gap.}
\begin{tabular}{cccccc}
\hline
\hline
Orientation&$T_{\rm c}$&$B_{\rm c}(0)$&$\alpha=\frac{\Delta}{k_{\rm B}T_{\rm c}}$&$\Delta$\\
           & (K)       &(mT)          & &(meV)\\
\hline
$B_{\rm ex}\parallel [100]$ (exp.\#1)&7.193(2)&79.828(4)&2.309(2) & 1.431(2)\\
$B_{\rm ex}\parallel [110]$ (exp.\#1)&7.191(2)&79.852(4)&2.311(2) & 1.432(2)\\
$B_{\rm ex}\parallel [100]$ (exp.\#2)&7.190(2)&79.847(4)&2.312(2) & 1.432(2)\\
$B_{\rm ex}\parallel [111]$ (exp.\#2)&7.190(2)&79.871(4)&2.315(2) & 1.434(2)\\
\hline
Averaged                           &7.191(2)&79.850(4)&2.312(2)&1.432(2)\\
\hline
\hline
\end{tabular}
\end{table}

The parameters obtained from the analysis of the measured $B_{\rm c}(T)$ dependences by means of Eq.~\ref{eq:alpha-model_Padamsee} are summarized in Table~\ref{Table1}.
$\lambda_{\rm el-ph}=1.5$ was used, in accordance with $\lambda_{\rm el-ph}=1.48-1.55$ of elemental Pb reported in the literature \cite{Dynes_SSC_1972, Dynes_PPRB_1975, Allen_PRB_1987, Carbotte_RMP_1990}. The solid lines in Figs.~\ref{fig:Bc}~(a) and (b) are the fit of Eq.~\ref{eq:alpha-model_Padamsee} to the $B_{\rm c}(T)$ obtained in exp.\#2 with $B_{\rm ex}\parallel [100]$ (fit curves strongly overlap with each other, so only a single curve is shown, as a representative manner). The fitting curves for each individual experiment are presented in the Supplemental Material \cite{SupplementalMaterial}.

From the results presented in Fig.~\ref{fig:Bc} and Table~\ref{Table1}, the following three important points emerge:\\
(i) The values of the transition temperature $T_{\rm c}=7.191(2)$~K and the thermodynamic critical field $B_{\rm c}(0)=79.850(4)$~mT stay in agreement with the values reported in literature for high quality Pb single crystal samples \cite{Chanin_PRB_1972, Brandt_ZhETF_1975, Matthias_RMP_1963}. The tiny, $\simeq2$~mK $T_{\rm c}$ change between the exp.\#1 and exp.\#2 could be caused by slightly different thermal contact between the sample and the cryostat's cold plate. Note that the change from exp.\#1 to exp.\#2 required regluing of the sample to the cold plate. The same reason explains the 0.019(6)~mT difference in $B_{\rm c}(0)$ values obtained in exp.\#1 and exp.\#2 with $B_{\rm ex}\parallel [100]$.\\
(ii) Experiments with the external magnetic field applied along the diagonal directions ([110] and [111]) show a slight increase of $B_{\rm c}(0)$ values as compared to $B_{\rm ex}\parallel [100]$ set of experiments. The corresponding difference is 0.024(6)~mT for both diagonal orientations. Bearing in mind that $B_{\rm c}(0)$ is a measure of the condensation energy (see Eq.~\ref{eq:condensation-energy}) and considering that the superconducting energy gap remains the same within experimental accuracy (see Table~\ref{Table1}), this would imply that the density of states at the Fermi levels depends on the crystallographic direction. The effect is, however, very small and accounts for $\sim 0.04-0.1$\% increase of $N_{\rm F}(0)$ along [110] and [111] directions. \\
(iii) The reduced gap, {\it i.e.}, the ratio of the superconducting energy gap to $T_{\rm c}$, does not depend on orientation. For all four measurements performed in three different field orientations (along [100], [110], and [111] crystal axes) the values of $\alpha=\Delta/k_{\rm B}T_{\rm c}$ stay the same within experimental uncertainty.  \\

The experimental data obtained in the present study are not consistent with the presence of two distinct superconducting energy gaps in elemental Pb for the two following reasons: \\
First of all, anisotropies of the Fermi surface sheets formed by two energy bands in elemental Pb are very much different. The so-called `inner' Fermi surface is almost spherical in shape, while the `outer' one has a tubular structure  \cite{Ruby_PRL_2015, Floris_PRB_2007, Anderson_PhyRev_1965}. For this reason, energy gaps associated to each Fermi surface are expected to have different contributions to the thermodynamic quantities, including $B_{\rm c}$, in various crystallographic  directions. Following discussions of Ref.~\onlinecite{Ruby_PRL_2015}, the contribution of the gap opened in the outer (tubular) Fermi surface needs to be highest along [110], intermediate along [100] and  smallest along the [111] crystal axes. Considering the fact that gaps opened in the inner and outer Fermi surfaces are different, this would imply the dependence of $B_{\rm c}(T)$, and so $D(T)$, on the crystallographic direction. Our experiments reveal, however, that all measured $B_{\rm c}(T)$ and $D(T)$ dependences are identical.\\
Second, within the two-gap scenario, the deviation curve $D(t^2)$ consists of the contribution from both energy gaps \cite{Khasanov_AuBe_PRB_2020}. The presence of two superconducting energy gaps in elemental Pb should let $D(t^2)$ stay somewhere in between, or even within, the $\alpha_{\rm small}$ and/or $\alpha_{\rm big}$ branches. The experimental $D(t^2)$ points lie, however, on the top of $\alpha_{\rm big}$ branch  (see Fig.~\ref{fig:Deviation-Tunneling}). Consequently, any admixture of the smaller gap to the measured $D(t^2)$ becomes impossible.

It is worth to emphasize, that the results of bulk $\mu$SR measurement presented here may not rule out the possibility that the superconducticting response of elemental Pb changes by approaching the surface.
Following predictions of the self-consistent two-gap model \cite{Suhl_PRL_1959, Bussmann-Holder_EPL_2004, Kogan_PRB_2009, Gupta_PRB_2020}, even though the superconducting gaps are opened at two different
Fermi surfaces, the temperature evolutions, as well as the absolute values of these gaps would stay the same in a case of strongly coupled electronic bands: $\Delta_1(T)=\Delta_2(T)=\Delta(T)$, $T_{\rm c1}=T_{\rm c2}=T_{\rm c}$. Here indexes `1' and `2' denote the first and the second gap, respectively. By decreasing the interband coupling, the gaps start to behave differently, but they are still opened at the same transition temperature: $\Delta_1(T)\neq\Delta_2(T)$, $T_{\rm c1}=T_{\rm c2}=T_{\rm c}$. Finally, in a case of zero coupling between the bands, the temperature dependencies of the gaps are different, and
the gaps vanish at two different temperatures: $\Delta_1(T)\neq\Delta_2(T)$, $T_{\rm c1}\neq T_{\rm c2}$. 

Based one the above presented arguments one may assume that the difference between our $\mu$SR measurements, where muons stopped at a distance of $\simeq0.23$~mm from the Pb sample surface (see the Supplemental Material \cite{SupplementalMaterial}) and tunneling experiments, which are sensitive to the top most atomic layer of the material, could be caused by the corresponding decrease of the interband coupling constant. The weakening of the coupling constant between two electronic bands by approaching the surface, would naturally explain the two-gap observation by means of tunneling. An interesting follow up study to consider, would be to perform low-energy muon experiments, which may allow probing down to $\sim10$~nm thick surface layer \cite{Morenzoni_LEM_JPCM_2004}.

To conclude, measurements of the temperature evolution of the thermodynamic critical field $B_{\rm c}$  along three crystallographic directions in elemental Pb were performed by means of muon-spin rotation/relaxation. Experiments which were carried out by applying the external magnetic field along the [100], [110], and [111] crystal axes suggest the presence of a {\it single} superconducting energy gap with the absolute value {\it independent} of the crystallographic orientation.  Overall, our experiments imply that elemental Pb is an isotropic single-gap superconductor.

\vspace{0.2cm}

The present work was performed at the Swiss Muon Source (S$\mu$S), Paul Scherrer Institute (PSI, Switzerland). The research work of R.G. was supported by the Swiss National Science Foundation (SNF-Grant No. 200021-175935).

\end{document}